\documentclass{article}
\usepackage{microtype}
\usepackage{graphicx}
\usepackage{subcaption}
\usepackage{booktabs}
\usepackage{array}
\usepackage{amsmath,amssymb}
\usepackage{xspace}
\usepackage{xcolor}
\usepackage{url}
\usepackage{xurl}
\usepackage{makecell}
\usepackage{needspace}
\usepackage{flafter}
\usepackage{placeins}
\usepackage{tikz}
\usepackage{pgfplots}
\pgfplotsset{compat=1.17}
\usetikzlibrary{positioning,arrows.meta,calc,patterns,patterns.meta,fit,backgrounds,decorations.pathreplacing,shapes.geometric}
\definecolor{rInk}{HTML}{262626}      
\definecolor{rFrame}{HTML}{7F7F7F}    
\definecolor{rGrey}{HTML}{7F7F7F}
\definecolor{rLight}{HTML}{E8E8E8}    
\definecolor{rMid}{HTML}{BFBFBF}      
\definecolor{rRestartF}{HTML}{F5D9BF} 
\definecolor{rRestart}{HTML}{B8743C}  
\definecolor{rRollF}{HTML}{D9E3BF}    
\definecolor{rRoll}{HTML}{6E8B3D}     
\definecolor{rResumeF}{HTML}{B7DCEF}  
\definecolor{rResume}{HTML}{3B5A8C}   
\definecolor{rGroundF}{HTML}{9CB4E9}  
\definecolor{rGround}{HTML}{5B6FA8}
\definecolor{rFailF}{HTML}{E8CDCA}    
\definecolor{rFail}{HTML}{A8382C}     
\tikzset{
  raf/.style={font=\sffamily\fontsize{7.4}{8.6}\selectfont, line cap=round, line join=round, >={Stealth[length=3pt,width=2.4pt]}, rInk},
  mcall/.style={circle, draw=rInk, line width=0.5pt, fill=white, minimum size=6.2pt, inner sep=0pt},
  tcall/.style={rectangle, draw=rInk, line width=0.5pt, fill=rMid, minimum size=5.4pt, inner sep=0pt},
  kept/.style={fill=rLight},
  redo/.style={pattern={Lines[angle=45,distance=2.2pt,line width=0.45pt]}, pattern color=#1},
  newwork/.style={fill=#1},
  failx/.style={rFail, line width=1.1pt},
  lbl/.style={font=\sffamily\fontsize{6.8}{7.8}\selectfont, rInk, inner sep=1pt},
  sub/.style={font=\sffamily\fontsize{6.4}{7.4}\selectfont, rGrey, inner sep=1pt},
  panel/.style={draw=rFrame, line width=0.5pt},
  flow/.style={->, line width=0.6pt, rInk},
  box/.style={draw=rInk, line width=0.6pt, align=center, font=\sffamily\fontsize{6.8}{7.8}\selectfont, inner sep=2pt, fill=white, minimum height=0.62cm},
  subcap/.style={font=\sffamily\fontsize{6.8}{7.8}\selectfont, rInk, align=center},
  ttl/.style={anchor=west, font=\sffamily\bfseries\fontsize{7.4}{8.6}\selectfont, rInk},
}

\usepackage{algorithm}
\usepackage{algorithmic}

\usepackage{hyperref}

\usepackage[accepted]{mlsys2025}
\makeatletter
\renewcommand{\Notice@String}{Preprint.}
\makeatother
\hypersetup{
  pdftitle={Repeating Work Can Lower Energy per Successful Goal in LLM Agents},
  pdfauthor={Deepak Panigrahy, Stephen Abkin, Aakash Tyagi, David Kebo Houngninou, Rabi Mahapatra},
  pdfsubject={LLM agent failure recovery},
  pdfkeywords={LLM agents, failure recovery, energy per successful goal, prefix caching}
}

\newcommand{\cacheTokSave}{\checkval{about 45}}
\newcommand{\cacheEnergySave}{\checkval{2 to 8}}

\newcommand{\idleBoardW}{50.2}

\newcommand{\nBaseTasks}{198}

\newcommand{\nStable}{58}

\newcommand{\nEoneTasks}{30}
\newcommand{\nHeldOut}{28}

\newcommand{\tripLlm}{18}
\newcommand{\tripTool}{13}

\providecommand{\sameOutcomeModel}{\pending{sameOutcomeModel}}

\providecommand{\lawCtxMax}{\pending{lawCtxMax}}

\providecommand{\sweepQuadD}{\pending{sweepQuadD}}
\providecommand{\sweepQuadDLo}{\pending{sweepQuadDLo}}
\providecommand{\sweepQuadDHi}{\pending{sweepQuadDHi}}

\providecommand{\etwobEarlySrWarmOk}{\pending{etwobEarlySrWarmOk}}
\providecommand{\etwobEarlyFrWarmOk}{\pending{etwobEarlyFrWarmOk}}

\providecommand{\latePointPct}{70}
\providecommand{\minLlmCalls}{8}
\providecommand{\minToolCalls}{5}

\providecommand{\sweepQwenCachedRatioLong}{\pending{sweepQwenCachedRatioLong}}

\providecommand{\cacheEnergySave}{}\renewcommand{\cacheEnergySave}{8}
\providecommand{\cacheEnergySaveHi}{}\renewcommand{\cacheEnergySaveHi}{10}
\providecommand{\cacheEnergySaveLo}{}\renewcommand{\cacheEnergySaveLo}{6}
\providecommand{\cacheTokSave}{}\renewcommand{\cacheTokSave}{86}
\providecommand{\cacheTokSaveHi}{}\renewcommand{\cacheTokSaveHi}{87}
\providecommand{\cacheTokSaveLo}{}\renewcommand{\cacheTokSaveLo}{85}

\providecommand{\earOkEar}{}\renewcommand{\earOkEar}{100 of 111}

\providecommand{\earSaveVsRestart}{}\renewcommand{\earSaveVsRestart}{22.0} 
\providecommand{\earSaveVsRestartHi}{}\renewcommand{\earSaveVsRestartHi}{31.8} 
\providecommand{\earSaveVsRestartLo}{}\renewcommand{\earSaveVsRestartLo}{11.9} 

\providecommand{\earVsBestFixed}{}\renewcommand{\earVsBestFixed}{3.2} 
\providecommand{\earVsBestFixedHi}{}\renewcommand{\earVsBestFixedHi}{6.6} 
\providecommand{\earVsBestFixedLo}{}\renewcommand{\earVsBestFixedLo}{-1.9} 
\providecommand{\earVsResume}{}\renewcommand{\earVsResume}{7.6} 
\providecommand{\earVsResumeHi}{}\renewcommand{\earVsResumeHi}{12.5} 
\providecommand{\earVsResumeLo}{}\renewcommand{\earVsResumeLo}{2.0} 

\providecommand{\eoneBzeroOk}{}\renewcommand{\eoneBzeroOk}{84 of 90}

\providecommand{\eoneCallsMax}{}\renewcommand{\eoneCallsMax}{31}
\providecommand{\eoneCallsMin}{}\renewcommand{\eoneCallsMin}{14}

\providecommand{\eoneEarlyFrSavePooled}{}\renewcommand{\eoneEarlyFrSavePooled}{22}
\providecommand{\eoneEarlyFrSavePooledHi}{}\renewcommand{\eoneEarlyFrSavePooledHi}{35}
\providecommand{\eoneEarlyFrSavePooledLo}{}\renewcommand{\eoneEarlyFrSavePooledLo}{9}

\providecommand{\eoneLateSrSavePooled}{}\renewcommand{\eoneLateSrSavePooled}{24}
\providecommand{\eoneLateSrSavePooledHi}{}\renewcommand{\eoneLateSrSavePooledHi}{36}
\providecommand{\eoneLateSrSavePooledLo}{}\renewcommand{\eoneLateSrSavePooledLo}{10}

\providecommand{\ethreeCtrlVsFrWarmSave}{}\renewcommand{\ethreeCtrlVsFrWarmSave}{13.3} 

\providecommand{\idleBoardW}{}\renewcommand{\idleBoardW}{47.8}

\providecommand{\latePointPct}{}\renewcommand{\latePointPct}{70} 

\providecommand{\minLlmCalls}{}\renewcommand{\minLlmCalls}{8}
\providecommand{\minToolCalls}{}\renewcommand{\minToolCalls}{5}

\providecommand{\nBaseTasks}{}\renewcommand{\nBaseTasks}{198}

\providecommand{\nEoneTasks}{}\renewcommand{\nEoneTasks}{30}

\providecommand{\nHeldOut}{}\renewcommand{\nHeldOut}{28}

\providecommand{\nStable}{}\renewcommand{\nStable}{58}

\providecommand{\sameOutcomeModel}{}\renewcommand{\sameOutcomeModel}{98.5} 

\providecommand{\timeEpgFrWarmEarly}{}\renewcommand{\timeEpgFrWarmEarly}{20.3} 
\providecommand{\timeEpgFrWarmLate}{}\renewcommand{\timeEpgFrWarmLate}{27.7} 
\providecommand{\timeEpgSrWarmEarly}{}\renewcommand{\timeEpgSrWarmEarly}{25.9} 
\providecommand{\timeEpgSrWarmLate}{}\renewcommand{\timeEpgSrWarmLate}{21.3} 
\providecommand{\tripLlm}{}\renewcommand{\tripLlm}{18}
\providecommand{\tripTool}{}\renewcommand{\tripTool}{13}

\providecommand{\xstarRuleOnlineVsFr}{}\renewcommand{\xstarRuleOnlineVsFr}{16.0} 

\providecommand{\ctrlContextChangedDecisionsN}{}\renewcommand{\ctrlContextChangedDecisionsN}{44 of 111} 
\providecommand{\etwobEarlyFrWarmOk}{}\renewcommand{\etwobEarlyFrWarmOk}{12 of 13} 
\providecommand{\etwobEarlySrWarmOk}{}\renewcommand{\etwobEarlySrWarmOk}{3 of 13} 
\providecommand{\lawCtxMax}{}\renewcommand{\lawCtxMax}{7{,}333} 
\providecommand{\sweepQuadD}{}\renewcommand{\sweepQuadD}{0.546} 
\providecommand{\sweepQuadDHi}{}\renewcommand{\sweepQuadDHi}{0.584} 
\providecommand{\sweepQuadDLo}{}\renewcommand{\sweepQuadDLo}{0.508} 
\providecommand{\sweepQwenCachedCtxMax}{}\renewcommand{\sweepQwenCachedCtxMax}{32{,}672} 
\providecommand{\sweepQwenCachedLinB}{}\renewcommand{\sweepQwenCachedLinB}{0.0023} 
\providecommand{\sweepQwenCachedNCalls}{}\renewcommand{\sweepQwenCachedNCalls}{60} 
\providecommand{\sweepQwenCachedRatioFourK}{}\renewcommand{\sweepQwenCachedRatioFourK}{0.54} 
\providecommand{\sweepQwenCachedRatioLong}{}\renewcommand{\sweepQwenCachedRatioLong}{0.10} 
\providecommand{\sweepQwenUncachedNCalls}{}\renewcommand{\sweepQwenUncachedNCalls}{60} 
\providecommand{\sweepQwenUncachedQuadD}{}\renewcommand{\sweepQwenUncachedQuadD}{7.81e-07} 
\providecommand{\sweepQwenUncachedQuadDHi}{}\renewcommand{\sweepQwenUncachedQuadDHi}{8.74e-07} 
\providecommand{\sweepQwenUncachedQuadDLo}{}\renewcommand{\sweepQwenUncachedQuadDLo}{7.01e-07} 

\newcommand{\qEarlyFrOk}{112 of 120}  
\newcommand{\qEarlySrOk}{91 of 120}

\newcommand{\qLateToolRbOk}{29 of 30} 
\newcommand{\qLateToolSrOk}{24 of 30} 
\newcommand{\qLateToolSgOk}{26 of 30} 
\newcommand{\misEarlyFrOk}{31 of 32}  \newcommand{\misEarlySrOk}{26 of 32}
\newcommand{\gptEarlyFrOk}{43 of 48}  \newcommand{\gptEarlySrOk}{29 of 48}
\newcommand{\pilotEarlyFrOk}{40 of 40} \newcommand{\pilotEarlySrOk}{24 of 40}
\newcommand{\sampEarlyFrOk}{58 of 60} \newcommand{\sampEarlySrOk}{47 of 60}

\newcommand{\trRbWarm}{98} \newcommand{\trSrWarm}{13} \newcommand{\trSrWarmJson}{12} \newcommand{\trPredGap}{69}
\newcommand{\earSaveVsRestartInt}{22} \newcommand{\earVsResumeInt}{8}

\newcommand{\sameHistMain}{238} \newcommand{\sameHistRb}{240} \newcommand{\sameHistCross}{239}
\newcommand{\gapWarmS}{0.15} \newcommand{\gapColdS}{0.17} \newcommand{\gapShareMed}{1} \newcommand{\gapShareMax}{2}
\newcommand{\rbBatchRatioPct}{5}
\newcommand{\rbFrEarlyTool}{2.9} \newcommand{\rbFrEarlyToolLo}{$-$9.7} \newcommand{\rbFrEarlyToolHi}{13.2}   
\newcommand{\rbSrEarlyTool}{22.5} \newcommand{\rbSrEarlyToolLo}{1.1} \newcommand{\rbSrEarlyToolHi}{42.6}     
\newcommand{\rbFrLateTool}{20.2} \newcommand{\rbFrLateToolLo}{6.8} \newcommand{\rbFrLateToolHi}{31.2}       
\newcommand{\rbSrLateTool}{3.6} \newcommand{\rbSrLateToolLo}{$-$21.0} \newcommand{\rbSrLateToolHi}{24.5}     
  
\newcommand{\rbWarmColdHeld}{3.0} \newcommand{\rbWarmColdHeldLo}{$-$3.7} \newcommand{\rbWarmColdHeldHi}{8.4}
\newcommand{\cachedSharePct}{91}  

\renewcommand{\earVsBestFixedLo}{$-$1.9}

\definecolor{insightbg}{HTML}{ECECEC}
\newcommand{\finding}[2]{\par\smallskip\noindent\colorbox{insightbg}{\parbox{\dimexpr\columnwidth-2\fboxsep\relax}{\textbf{Finding #1.} #2}}\par\smallskip}


\AtBeginDocument{\clubpenalty=10000 \widowpenalty=10000 \displaywidowpenalty=10000}

\mlsystitlerunning{Repeating Work Can Lower Energy per Successful Goal in LLM Agents}

\begin{document}

\twocolumn[
\mlsystitle{\texorpdfstring{Repeating Work Can Lower Energy\\per Successful Goal in LLM Agents}{Repeating Work Can Lower Energy per Successful Goal in LLM Agents}}

\begin{mlsysauthorlist}
\mlsysauthor{Deepak Panigrahy}{tamu}
\mlsysauthor{Stephen Abkin}{tamu}
\mlsysauthor{Aakash Tyagi}{tamu}
\mlsysauthor{David Kebo Houngninou}{tamu}
\mlsysauthor{Rabi Mahapatra}{tamu}
\end{mlsysauthorlist}

\mlsysaffiliation{tamu}{Texas A\&M University, College Station, TX, USA}

\mlsyscorrespondingauthor{Deepak Panigrahy}{deepakp@tamu.edu}

\mlsyskeywords{LLM agents, failure recovery, energy per successful goal, prefix caching}

\vskip 0.3in

\begin{abstract}
When a temporary call failure interrupts a tool-using LLM agent, the framework must choose how much completed work to repeat.
Repeating less completed work intuitively saves computation, but recovery mechanisms also change subsequent execution and the chance of success.
We compare recovery mechanisms by energy per successful goal (EpG), which divides the energy of all task runs (including failed runs) by the number that finish correctly.
In controlled vLLM experiments on one node, we measure board energy above idle for eight recovery options on reliably solved multi-turn tasks, with selected comparisons across three model families.
For the model tested across all eight options, with the prefix cache retained, starting over has \eoneEarlyFrSavePooled{} percent lower EpG than continuing after early failures. 
Under the same cache condition, continuing has \eoneLateSrSavePooled{} percent lower EpG than starting over after late failures.
Among all eight recovery options, repeating only the interrupted turn has the lowest observed EpG after early and late tool failures on this model.
Holding recovery fixed, retaining the prefix cache reduces prompt tokens needing fresh computation by \cacheTokSave{} percent but reduces recovery energy by only \cacheEnergySave{} percent at contexts up to \lawCtxMax{} tokens.
These measurements and a condition relating recovery energy and success show when repeating work lowers EpG, helping framework developers choose how their agents recover on their intended models and tasks.

\end{abstract}
]

\printAffiliationsAndNotice{}

\section{Introduction}
\label{sec:intro}

Tool-using LLM agents manage files and book travel through sequences of model and tool calls, so a failed call can interrupt a task after substantial work.
We study transient failures (i.e., temporary failures of a model or tool call that clear so execution can continue).
The framework running the agent must then decide how much completed work to repeat.
Repeating less work avoids immediate recomputation, but the recovery also changes how execution continues and whether the task finishes correctly.
Which recovery uses the least energy for a correct result?

\begin{figure*}[t]
\centering
\includegraphics[width=\textwidth]{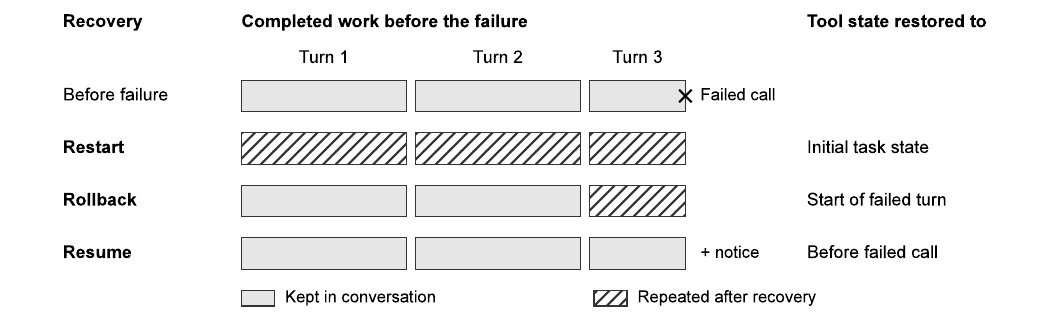}
\caption{\textbf{What each recovery keeps and repeats.}
A failure interrupts turn 3. The blocks represent work completed before the failure, grouped by user turn, with a partial block for turn 3. Gray blocks remain in the conversation, and hatched blocks show work to repeat. The right column identifies the earlier state to which the tools are restored. Resume adds an error notice. Grounded resume also includes the tools' actual state in that notice. We separately retain (warm) or clear (cold) the prefix cache for each recovery.}
\label{fig:concept}
\end{figure*}

Figure~\ref{fig:concept} compares recoveries by the conversation they retain and the tool state they restore (i.e., the data maintained by the tools).
Restarting begins the task again from the first user message and resets the tools to their initial state.
To repeat only the interrupted turn (i.e., the calls serving one user message), rollback restores both conversation and tools to that turn's beginning.
Resuming keeps the conversation and the current tool state and continues with an error notice~\cite{wu2024autogen,langgraph}; for the clean failures, the failed call leaves the tools as they were just before it.
Grounded resume adds the actual state of the tools touched by the interrupted turn to that notice.

Resuming preserves completed work, and the server can also reuse model state already computed for the beginning of a prompt.
A prefix cache stores this state, so keeping the cache warm (i.e., retaining the computed state for reuse during recovery) avoids processing much of the earlier conversation again~\cite{openaipromptcaching,anthropicpromptcaching}.
Retaining the conversation also affects what the agent does next, and a failed step kept in that conversation can lead to repeated calls or incorrect results~\cite{whyretrying2026,backfires2026}.
For a developer deciding how agents should recover, preserving completed work is useful when it lowers the energy needed for correct task completion.

To compare recoveries on that basis, we include the energy spent on unsuccessful runs.
A goal is one task run, including its initial execution and any recovery.
Mean energy per run divides total energy by the number of runs, while energy per successful goal (EpG) divides the same total by the number that finish correctly~\cite{panigrahy2026epg}.
A recovery that uses less energy per run can still require more energy per successful goal if it succeeds less often.
Table~\ref{tab:glance} illustrates this difference after a late tool error. Resuming uses the least energy per run, while rollback completes more runs correctly than resuming and has the lowest energy per successful goal.

\begin{table}[htbp]
\centering
\caption{\textbf{Energy per run and energy per successful goal favor different recoveries.} Qwen3 30B, 30 tasks, each interrupted by one late tool error. Each run includes its recovery. Mean run energy averages over successful and failed runs. Energy per successful goal (EpG) divides their total energy by the number that finish correctly. Grounded resume adds the tools' actual state to the error notice supplied when resuming. All four recoveries use a warm prefix cache. Bold marks the lowest observed EpG among these recoveries.}
\label{tab:glance}
\vskip 0.1in
\footnotesize
\setlength{\tabcolsep}{4pt}
\begin{tabular}{@{}lcccc@{}}
\toprule
 & Restart & Resume & \makecell{Grounded\\resume} & Rollback \\
\midrule
Correct runs & 28/30 & 24/30 & 26/30 & 27/30 \\
Mean run energy, J & 1{,}711 & 1{,}208 & 1{,}337 & 1{,}295 \\
\makecell{Energy per successful\\goal (EpG), J} & 1{,}833 & 1{,}510 & 1{,}543 & \textbf{1{,}439} \\
\bottomrule
\end{tabular}
\end{table}

\paragraph{Motivation.}
In production telemetry from thousands of organizations, about one in twenty model requests failed in February 2026~\cite{datadog2026aieng}.
A framework's choice of recovery is reused across tasks and agents~\cite{wu2024autogen,langgraph}, so its energy consequences accumulate over repeated failures.
Developers can change how much work recovery repeats while keeping the model and hardware fixed.

\paragraph{Goal.}
We seek to determine when repeating completed work reduces the energy needed for correct task completion.
Recovery research develops ways to detect failures, restore execution state, and choose a recovery point, and evaluates whether the recovered task finishes correctly~\cite{rir2026,zhang2026recoverability,agentrewind2026,dubey2026realtimedetectionrepairllm}.
Serving research manages prefix caches to reduce latency, price, or carbon cost~\cite{abhyankar2024infercept,pan2025kvflow,khailo2026keepingcachewarmpays,tian2026cachepromptitsgreen}, and agent energy studies characterize where an agent's energy goes~\cite{whereagentslive2026,blackwellagents2026}.
The missing comparison is how these recovery choices affect measured energy per successful goal when both prefix-cache reuse and unsuccessful runs are included.
To address this gap, we compare the four recovery operations with the prefix cache retained or cleared, giving eight options.
We test them after matched failures (i.e., failures of the same type injected at the same planned call number).
Moreover, we ask four research questions in which we answer in Section~\ref{sec:results}:
\begin{description}
\item[RQ1:] When does restarting the task use less energy per successful goal than resuming from the failed step?
\item[RQ2:] How does repeating only the interrupted turn compare with restarting the task or resuming from the failed step?
\item[RQ3:] How much recovery energy does a warm prefix cache save when the recovery is held fixed?
\begin{samepage}
\item[RQ4:] Can information available when a failure occurs guide recovery choices that reduce energy per successful goal on new tasks?
\par\end{samepage}
\end{description}

We select multi-turn tasks from the Berkeley Function Calling Leaderboard (BFCL)~\cite{patil2025bfcl} that the model completes reliably without injected failures.
Its checker determines whether a run succeeds by comparing the tools' final states after each turn with the expected states.
To answer RQ1, we inject one transient failure early or late in each run and compare restarting with resuming.
We add rollback to answer RQ2, and we retain or clear the prefix cache for each recovery to answer RQ3.
We test all eight options on Qwen3 30B, our main model, across 30 tasks (the main study), and selected options on two other model families.
To answer RQ4, we fit decision rules on earlier experiments and execute the recoveries they choose as agents run 28 additional tasks excluded from fitting.
We serve the models with vLLM~\cite{kwon2023vllm} and measure energy at the board of one GB10 node while running one goal at a time, using its power with the model loaded and no requests as the idle baseline.

\paragraph{Summary of findings.}
In the main study, restarting uses less energy per successful goal than resuming after early failures, while resuming uses less than restarting after late failures (\S\ref{sec:point}).
Restarting's early advantage reflects its higher success rate and the energy resuming uses above the estimated energy of unfinished work. Our cost model weighs these contributions against the energy already spent, which is larger after late failures.
After early failures, resuming succeeds less often across the three tested model families, with the energy ranking also varying by model and failure type.
After tool failures in the main study, repeating only the interrupted turn reduces energy per successful goal relative to resuming early and restarting late (\S\ref{sec:drop}).
Across the main study's failure types and points, warm rollback has the smallest mean excess cost among the fixed recoveries tested.
At prompt lengths up to \lawCtxMax{} tokens, a warm prefix cache reduces input tokens requiring fresh computation by \cacheTokSave{} percent but reduces energy used after failure by only \cacheEnergySave{} percent (\S\ref{sec:cache}).
On tasks excluded from fitting, the token rule using the failure type and the number of calls already made reduces energy per successful goal relative to always resuming and always restarting (\S\ref{sec:choose}).
It mostly chooses rollback with a warm prefix cache, and its estimated saving over always rolling back cold is within the variation across tasks.
We propose that developers choose how their agents recover by measuring energy per successful goal for the models and tasks those agents will run.

\paragraph{Contributions.}
\begin{itemize}
\item We characterize the energy per successful goal of eight recovery options and show when repeating work lowers it, using 30 multi-turn tasks with selected comparisons across three model families (\S\ref{sec:method}, \S\ref{sec:results}).
\item We build a setup for measuring recovery energy against verified task outcomes and release the recovery decision library~\cite{rafcode2026} (\S\ref{sec:method}).
\item We derive a condition for when restarting lowers energy per successful goal and evaluate its terms from measured executions (\S\ref{sec:metric}, \S\ref{sec:point}).
\end{itemize}

\section{Background and Related Work}
\label{sec:background}

\paragraph{Agent execution and call failures.}
A tool-using agent completes a task by asking a model what to do next, executing the requested tools, and adding their results to the conversation~\cite{yao2023react,schick2023toolformer,patil2024gorilla}.
Each model call receives the conversation so far as its input prompt, which grows as replies and tool results are added.
In the road-trip task of our study, six user requests take the agent from looking up zip codes to starting the engine.
Completing those requests requires \tripLlm{} model calls and \tripTool{} tool calls.

Because execution depends on both model responses and tool results, a failure on either side interrupts the agent's progress.
We study four call failures (Section~\ref{sec:method}).
On the model side, a request is rejected by a rate limit or the reply is malformed so the framework cannot parse its tool call.
On the tool side, a call returns an error or reaches its timeout without a response.
Automatic retries by a model client or tool library may resolve a failure before the agent needs to recover.
Our comparison concerns failures passed to the agent framework, where it must choose between continuing from retained progress and repeating earlier work.

\paragraph{Restoring execution after a failure.}
Checkpointing saves a program's execution state so it can return to that state after a failure.
Classical models choose how often to save state by balancing the cost of saving it against the work lost when failures occur~\cite{young1974checkpoint,daly2006checkpoint}.
For an agent, recovery also determines which conversation messages and tool data will guide its subsequent actions.
Rollback-Induced Reflection restores an earlier state while retaining lessons from the failed execution~\cite{rir2026}.
\citet{zhang2026recoverability} check the conditions for reusing saved progress before execution continues.
AgentRewind restores saved versions of both the agent's conversation and its workspace, and tests their roles by disabling each restoration separately~\cite{agentrewind2026}.
Our rollback also restores conversation and tool data together, so its measured energy and success describe the combined recovery operation.

After execution is restored, the model's next decision still depends on the messages remaining in the conversation.
\citet{whyretrying2026} analyzes retries under the assumption that a failed attempt left in the conversation increases the chance of further errors.
In experiments with small models, \citet{backfires2026} finds that showing a failed call and its error message increases the likelihood of repeating that call.
For agents with persistent memory, \citet{wang2026faultymemoriescorrectedactions} trace how faulty memories affect later decisions and selectively repeat affected work while preserving information supported by independent evidence.
We compare recovery choices by measured board energy per successful goal, varying cache retention to account for the prompt computation the server reuses.

\paragraph{Reusing prompt computation during recovery.}
A model call processes its input prompt during prefill and then generates the reply one token at a time during decoding~\cite{zhong2024distserve,patel2024splitwise}.
With prefix caching, an inference engine such as vLLM reuses computed state for the beginning of a matching prompt, reducing the work needed for prefill~\cite{kwon2023vllm,zheng2024sglang,vllmprefixcaching}.
Cached input tokens reuse this state, while uncached prompt tokens require fresh prefill.
Output tokens are generated during decoding, and context length is the number of input tokens supplied to a model call.
A warm recovery retains the prefix cache, while a cold recovery starts after it has been cleared.
Both supply the conversation specified by the recovery operation, so retaining conversation history and reusing its computed state are separate choices.

Serving systems manage prefix caches according to their expected reuse to reduce latency~\cite{abhyankar2024infercept,luo2025autellix,pan2025kvflow,li2025continuum,kang2026thunderagent}.
Other studies evaluate the provider charges for keeping a prompt cached through tool pauses~\cite{khailo2026keepingcachewarmpays}, or choose what to cache according to carbon cost, as in GreenCache~\cite{tian2026cachepromptitsgreen}.
The energy of a model call also depends on model size, output length, batch size, context length, and inference phase~\cite{samsi2023words,luccioni2024power,chung2025mlenergy}.
\citet{prefilldecode2026} analyze how prompt processing, reply generation, context length, and batching affect energy.
Measured recovery energy covers input processing, output generation, tool execution, and waits, so reusing prompt computation saves only part of that energy.
We measure that contribution by keeping the recovery operation fixed and comparing its retained-cache and cleared-cache versions.

\paragraph{Energy and verified task completion.}
Evaluating accuracy and cost together shows what an agent spends to produce useful results~\cite{kapoor2025agents}.
Energy per successful goal includes the energy of failed runs in the total charged to correct completions~\cite{panigrahy2026epg}.
OSWorld-Human follows a related principle for computer-use agents, accounting for unnecessary steps and work spent on failed tasks~\cite{abhyankar2026osworldhuman}.
It identifies rollback and prefix caching as promising ways to improve efficiency.
We measure how these choices affect energy per successful goal after controlled call failures.

Other approaches stop agents likely to fail~\cite{agentstop2026,doomed2026,failfast2026}, detect failures from measurements collected during execution~\cite{dubey2026realtimedetectionrepairllm}, or measure tokens spent after an early warning of failure~\cite{failaware2026}.
At the infrastructure level, systems control GPU power and schedule model execution~\cite{you2023zeus,patel2024polca,chung2024perseus}, while fault-tolerant serving systems restore interrupted computation~\cite{strati2024dejavu,lumen2026}.
Comparing recoveries requires accounting for the energy already spent before failure and the energy of runs that finish incorrectly.
We next use this accounting to derive when repeating completed work reduces the energy needed for a correct result.

\section{Energy per Successful Goal}
\label{sec:metric}

For our BFCL tasks, the benchmark's automated checker determines success by comparing the tools' states after each user turn with the expected states.
An attempt is one execution segment, ending at the injected failure or when execution finishes.
A goal is one task run, including the initial attempt and any recovery attempts.
Goal energy sums the measured energy of those attempts, while recovery energy counts only the attempts after the initial failure.
We specify what is metered in Section~\ref{sec:method}.

Mean energy per run divides total goal energy by the number of runs, including successful and failed runs.
For a group with at least one correct goal, energy per successful goal (EpG) divides that same total by the number of correct goals, giving joules per correct completion~\cite{panigrahy2026epg}.
In Table~\ref{tab:glance}, warm resume averages 1{,}208~J per run and completes 24 of 30 goals correctly, giving approximately $1{,}208/(24/30)=1{,}510$~J per successful goal.
Warm rollback averages 1{,}295~J per run and completes 27 of 30 goals correctly, giving $1{,}295/(27/30)\approx1{,}439$~J per successful goal.
Rollback's higher success makes it cheaper per correct completion despite its higher energy per run.

\paragraph{When repeating work reduces energy.}
We now compare restarting with resuming from the same state at the failure.
Let $S$ be the energy already spent.
The experiments approximate this comparison with separate executions, checking agreement of their calls and replies before the failure (Section~\ref{sec:method}).
For a recovery $o$, let $R_o$ be its expected recovery energy, and let $p_o>0$ be its probability of completing the goal correctly.
This expectation averages the energy of both successful and failed recoveries.
The expected goal energy is $S+R_o$, so the energy per successful goal is
\begin{equation}
J_o = \frac{S + R_o}{p_o}.
\label{eq:J}
\end{equation}
Dividing by $p_o$ charges more of the pre-failure energy to each correct completion when recovery succeeds less often.

Let $E$ be the expected energy of a restart's recovery attempt, and let $p_R>0$ be its probability of completing the goal correctly.
Because restarting repeats the task, our simplified cost model uses $E-S$ to estimate the energy of unfinished work.
Resuming has expected recovery energy $E-S+\Delta$ and succeeds with probability $p_K>0$.
The residual $\Delta$, which we call additional recovery energy, is positive when resuming uses more energy than the estimate $E-S$.
Restarting's goal energy is $S+E$, while resuming's is $S+(E-S)+\Delta=E+\Delta$.
Substituting these totals into Equation~\ref{eq:J}, we find that restarting is cheaper exactly when $(S+E)/p_R<(E+\Delta)/p_K$.
Multiplying by the positive success probabilities and rearranging, we obtain
\begin{equation}
S < \Bigl(\frac{p_R}{p_K} - 1\Bigr) E + \frac{p_R}{p_K}\,\Delta .
\label{eq:rule}
\end{equation}
On the right, the success-rate term $(p_R/p_K-1)E$ accounts for the difference in success rates, while the additional-energy term $(p_R/p_K)\Delta$ weights the residual by $p_R/p_K$.
Restarting is cheaper when their sum exceeds the energy already spent, $S$.

We estimate $S$ by averaging the mean energies before failure for restart and resume, and estimate $E$ from mean recovery energy when restarting.
We compute $\Delta$ as the mean recovery energy when resuming minus $E-S$ (\hyperref[app:derivation]{Appendix~\ref*{app:derivation}}).
We use the same estimate of energy spent before failure for both recoveries. Attributing $\Delta$ to particular actions requires a separate comparison.

With equal success rates, Equation~\ref{eq:rule} reduces to $S<\Delta$.
Restarting then saves energy only when resuming uses more energy above the estimate than has already been spent.
With $\Delta=0$, a higher restart success rate can still make restarting cheaper when $S<(p_R/p_K-1)E$.
Success rates enter alongside the repeated computation studied in classical checkpoint models~\cite{young1974checkpoint,daly2006checkpoint}.
Holding the other terms fixed, a smaller $S$ favors restarting, which motivates comparing failures early and late in a task.

For rollback, let $T$ be the energy of the completed part of the interrupted turn that must be repeated, and let $p_B>0$ be rollback's success probability.
Assuming the remaining work has expected cost $E-S$, goal energy is $E+T$ and $J_B=(E+T)/p_B$ (\hyperref[app:derivation]{Appendix~\ref*{app:derivation}}).
We compare restarting with resuming after early and late failures (Section~\ref{sec:point}), then evaluate rollback's energy and correctness (Section~\ref{sec:drop}).

\section{Experimental Design}
\label{sec:method}

\begin{figure}[htbp]
\centering
\includegraphics[width=\columnwidth]{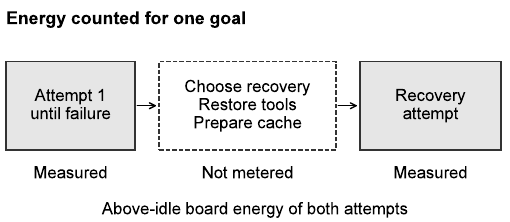}
\caption{\textbf{One goal includes the initial attempt and recovery.}
One planned call failure interrupts the first attempt. The gray boxes show the measured attempts, whose energies are summed for the goal. Choosing the recovery, restoring tools, and preparing the cache occur in the dashed box between attempts and are unmetered. Boxes represent execution stages, and arrows give their order.}
\label{fig:setup}
\end{figure}

We compare how much energy each recovery uses to complete a task correctly after the same planned failure.
Changing the recovery varies how much work the agent repeats, while retaining or clearing the prefix cache varies how much prompt computation the server reuses.
Early and late failures change how much work is already complete.
The benchmark checks correctness independently of the agent's report.
We measure the same parts of execution for each recovery and include energy from failed goals, supplying the terms in Equation~\ref{eq:rule} (Figure~\ref{fig:setup}).

\paragraph{Models and tasks.}
The main study uses Qwen3 30B A3B Instruct 2507 in FP8~\cite{qwen3}, served with vLLM~\cite{kwon2023vllm} and prefix caching enabled.
We check whether the recovery ordering recurs in other model families using Mistral Small 3.2 24B in 8~bit~\cite{mistral2025small32} and gpt-oss 20B in MXFP4~\cite{gptoss2025modelcard}.
Each model keeps its representation fixed across recovery comparisons and runs on tasks it completes reliably.
The additional families are tested with warm restart, warm resume, and cold rollback.
A Qwen3 4B pilot compares restart and resume on 10 tasks.

To study failures that disrupt otherwise solvable tasks, we select from the BFCL multi-turn base set~\cite{patil2025bfcl,yan2025bfcl}, whose expected tool states are known.
Screening requires three successful runs without injected failures and the same numbers of model calls and tool calls in those runs.
We also require at least \minLlmCalls{} model calls and \minToolCalls{} tool calls, leaving room to inject failures at an early and a late point.
Of \nBaseTasks{} base tasks, \nStable{} meet these conditions for the main model.
We use \nEoneTasks{} for the main comparison and reserve the other \nHeldOut{} for testing decision rules on additional tasks.
The selected tasks involve vehicle controls, files, trading, travel bookings, messaging, and ticket management, with \eoneCallsMin{} to \eoneCallsMax{} model and tool calls combined in a run without injected failures.
Separate reference runs without injected failures succeed in \eoneBzeroOk{} goals. These are distinct from the screening runs.
Study counts for each experiment are given in \hyperref[app:setup]{Appendix~\ref*{app:setup}}, \hyperref[tab:studies]{Table~\ref*{tab:studies}}.
Runs in the main study keep decoding settings fixed at temperature~0 and model seed~0.
We also sample responses at temperature~0.7 to check whether the recovery ordering persists when token selection varies.

\paragraph{Failures and recoveries.}
We distinguish model-side failures (a malformed reply or rate-limit rejection) from tool-side failures (a tool error or timeout).
An early failure occurs at the second model call or second tool call, according to the side affected.
For a late failure, we set the call number near \latePointPct{} percent of the corresponding model-call or tool-call count in the task's first failure-free screening run, before running the recovery comparisons.
The failure occurs once, during the initial attempt.
A goal enters a comparison when its log confirms that the planned failure occurred once at its target (\hyperref[app:setup]{Appendix~\ref*{app:setup}}).
Each goal allows one recovery attempt, and each attempt has a limit of 40 model calls.

The main study compares restart, resume, rollback, and grounded resume, each with the prefix cache retained or cleared.
We randomize the order within two measurement batches~\cite{mytkowicz2009wrong}, first comparing restart with resume, then rollback with grounded resume on the same tasks and failure plan.
The four failure types at two positions give eight failure conditions per task.
With four operations and two cache states, this gives 1,920 goals with injected failures across the 30 main-study tasks, plus 90 reference goals without injected failures.
Each recovery has a separate initial execution, so we check how closely those executions agree before failure.
We replace generated call and request identifiers with common labels before comparing the calls and replies.
Within the restart/resume batch, these histories agree in \sameHistMain{} of 240 task/failure/position combinations. Within the rollback/grounded-resume batch, they agree in \sameHistRb{} of 240 combinations (\hyperref[app:setup]{Appendix~\ref*{app:setup}}).

The two model-side failure types give the same correctness outcome in \sameOutcomeModel{} percent of paired goals, and the two tool-side types agree in all paired goals.
We report grouped results by model side or tool side, pooling goals across the two types where stated.
A separate study examines wrong actions that a tool accepts, extending the comparison to misleading steps that leave no error notice.

\paragraph{Energy measurement.}
A-LEMS~\cite{panigrahy2026epg} measures power at the node's DC input, covering memory and power conversion as well as the GPU~\cite{blackwellagents2026}.
We sample board and chip-package power at 10~Hz and GPU power at 1~Hz.
The GB10 node pairs a Grace CPU and a Blackwell GPU in 128~GB of shared memory~\cite{nvidiagb10}.
We serve one goal at a time, so each attempt window holds only that goal's work, including a rate limit's wait.
We subtract the power drawn with the model loaded and no requests (\idleBoardW{}~W for the first batch, remeasured for the second) to measure energy above the idle baseline.
We measure each initial and recovery attempt, including waits within an attempt.
Choosing a recovery, restoring tools, and preparing the cache occur between these windows and are unmetered (Figure~\ref{fig:setup}).
To quantify the excluded intervals, the main study records median gaps of \gapWarmS{}~s with a warm cache and \gapColdS{}~s with a cold one, about \gapShareMed{} percent of a goal's attempt time and nearly equal across recoveries.
If these intervals drew the same mean above-idle power as the attempts, they would add about \gapShareMed{} percent to measured goal energy at the median and less than \gapShareMax{} percent for 95 percent of main-study goals.

\paragraph{Pairing and uncertainty.}
We calculate each percentage saving from the two recoveries' pooled EpGs on the tasks included in both.
We obtain 95\% confidence intervals (CIs) from 2,000 resamples of whole tasks with replacement, using the same task draw for both recoveries and all included failure conditions~\cite{efron1994bootstrap}.
This retains the dependence among outcomes from the same task.
Tables report EpG for their stated populations. Held-out savings use 27 tasks with all four failure conditions (108 goals per policy), while \hyperref[app:heldout]{Appendix~\ref*{app:heldout}}, \hyperref[tab:e3]{Table~\ref*{tab:e3}} reports all 111 goals per policy.
For the cache comparison, we instead calculate the reduction in pooled recovery energy or uncached prompt tokens and resample tasks in the same way.
An observed minimum identifies the cheapest recovery in that sample.
We describe a saving as beyond the variation across tasks when its paired interval excludes zero.

\paragraph{Decision rules.}
The main comparisons identify low-energy recoveries after observing their outcomes.
We next test whether those measurements support choosing a recovery when the failure occurs, using rules fitted before evaluation on the \nHeldOut{} held-out tasks.
The threshold rule uses the failure type and total number of model and tool calls already made, restarting below a fitted cutoff and resuming above it.
The token rule uses the same information to predict each option's uncached prompt tokens after failure and selects the smallest prediction.
Each option has its own fit, reflecting the input needed across the calls generated during recovery.
The learning controller updates distributions of success probabilities from observed outcomes.
Using Thompson sampling~\cite{thompson1933}, it samples success rates and chooses the smallest predicted recovery energy divided by the sampled rate.
This score uses recovery energy, while Equation~\ref{eq:J} also includes energy spent before failure.
Below its fitted cutoff, its available actions were limited to restart (\hyperref[app:heldout]{Appendix~\ref*{app:heldout}}).

\begin{figure*}[t]
\centering
\includegraphics[width=\textwidth]{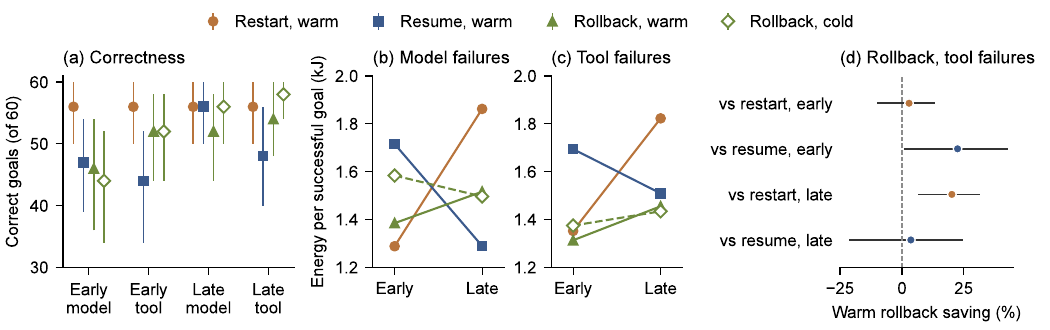}
\caption{\textbf{Restarting is cheaper early, resuming is cheaper late, and rollback has the lowest observed EpG after tool failures.}
Main study, Qwen3 30B. (a) Correct goals out of 60 for each failure side and position, with 95\% bootstrap intervals over the 30 tasks. Each task contributes two failure types. (b,c) Arithmetic means of the two failure-type EpG values within each side, in kJ measured at the board. Lines connect the two measured positions. (d) Paired saving in EpG of warm rollback over warm restart and warm resume after tool failures, with 95\% bootstrap intervals over tasks. Positive values favor rollback.}
\label{fig:costs}
\end{figure*}

We compare the three rules with applying warm restart, warm resume, or cold rollback at every failure, and use recovery policy for either a rule or one of these fixed choices.
For the main model, each policy executes the recovery it selects after early and late malformed replies and tool errors on the held-out tasks, giving 111 decisions per policy.
The threshold rule and learning controller use an earlier calibration detailed in \hyperref[app:heldout]{Appendix~\ref*{app:heldout}}.
One task's late tool-error case is excluded because its injection missed the planned call.

\section{Results}
\label{sec:results}

We compare how much work recovery should repeat, how much energy cache retention saves, and whether information available at failure guides a choice on new tasks.
Results are for the main study unless stated, and a positive saving means lower energy per successful goal for the first recovery named.
The recoveries reported in \hyperref[app:cells]{Appendix~\ref*{app:cells}}, \hyperref[tab:cells]{Table~\ref*{tab:cells}} cost 1.2 to 2.0~kJ per successful goal.

\subsection{RQ1: When Repeating Work Lowers Energy per Successful Goal}
\label{sec:point}

\begin{figure}[htbp]
\centering
\includegraphics[width=\columnwidth]{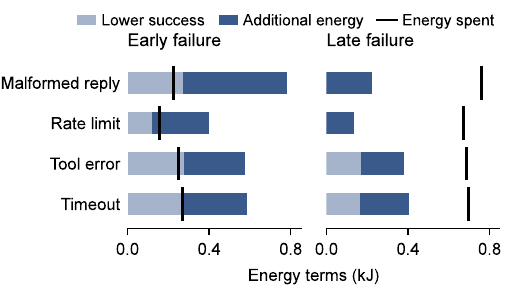}
\caption{\textbf{Lower success and additional recovery energy account for the early resume's cost.}
Main study, warm restart against warm resume, for each failure type early (left) and late (right). Bars stack the two terms on the right of Equation~\ref{eq:rule}, the cost of lower success (light) and additional recovery energy (dark), measured relative to the estimated unfinished work and weighted by $p_R/p_K$. The black line marks the energy already spent at the failure. Restarting lowers EpG where the bar exceeds the line.}
\label{fig:resumecost}
\end{figure}

Restarting lowers energy per successful goal after early failures and resuming lowers it after late ones, both beyond the variation across tasks (Figure~\ref{fig:costs}).
After an early failure, restarting repeats little completed work, making it easier for a higher success rate to offset that repetition.
With the cache warm, restarting completes \qEarlyFrOk{} goals correctly and resuming completes \qEarlySrOk{} goals correctly after early failures (Figure~\ref{fig:costs}a).
Restarting matches the failure-free reference success rate.
Restarting uses \eoneEarlyFrSavePooled{} percent less energy per successful goal than resuming (95\% CI \eoneEarlyFrSavePooledLo{} to \eoneEarlyFrSavePooledHi{} percent).
Equation~\ref{eq:rule} accounts for this difference through the success rates and the energy used after failure (Figure~\ref{fig:resumecost}, \hyperref[app:derivation]{Appendix~\ref*{app:derivation}}).
Across failure types, its success-rate term is 120 to 280~J and its additional-energy term is 280 to 510~J.
Their sum exceeds the 160 to 270~J already spent, satisfying the condition for restarting to use less energy per successful goal.

After late failures, the energy already spent is 670 to 760~J, exceeding the 140 to 400~J sum of the two terms (Figure~\ref{fig:resumecost}).
A late restart repeats most of the task, while after model-side failures both resume and restart complete 56 of 60 goals correctly.
Resuming uses \eoneLateSrSavePooled{} percent less energy per successful goal than restarting after late failures (95\% CI \eoneLateSrSavePooledLo{} to \eoneLateSrSavePooledHi{} percent).
After late tool failures, restarting still succeeds more often, but the success advantage is too small to offset repeating the work.
Time per successful goal follows the same early--late ordering (\hyperref[app:cells]{Appendix~\ref*{app:cells}}).

\paragraph{Correctness across models and sampled decoding.}
Restarting's early correctness advantage also appears under sampled decoding and on the additional models.
At temperature~0.7, resume completes \sampEarlySrOk{} goals correctly and restart completes \sampEarlyFrOk{} goals correctly.
For Mistral, resume completes \misEarlySrOk{} goals correctly and restart completes \misEarlyFrOk{} goals correctly. For gpt-oss, resume completes \gptEarlySrOk{} goals correctly and restart completes \gptEarlyFrOk{} goals correctly.
For Qwen3 4B, resume completes \pilotEarlySrOk{} goals correctly and restart completes \pilotEarlyFrOk{} goals correctly.
For Mistral after early tool failures, resuming fails one of eight goals for each failure type, and its energy per successful goal is lower than restarting's (\hyperref[app:cells]{Appendix~\ref*{app:cells}}, \hyperref[tab:m23]{Table~\ref*{tab:m23}}).
For these tasks and failures, restarting's higher success rate is too small to offset its higher measured goal energy.
In a separate study of wrong steps that tools accept without an error notice, excluding self-corrected cases leaves a different population for each recovery. Warm resume completes \etwobEarlySrWarmOk{} early goals correctly and warm restart completes \etwobEarlyFrWarmOk{} early goals correctly in their respective retained populations (\hyperref[app:setup]{Appendix~\ref*{app:setup}}).

\finding{1}{On the main model, restarting uses \eoneEarlyFrSavePooled{} percent less energy per successful goal than resuming after early failures. The resume's lower observed success rate and additional recovery energy account for this gap. After late failures, the energy already spent exceeds those two terms together, and resuming uses \eoneLateSrSavePooled{} percent less energy per successful goal than restarting (Figures~\ref{fig:costs} and~\ref{fig:resumecost}).}

\subsection{RQ2: How Rollback Compares with Restart and Resume}
\label{sec:drop}

After tool failures on the main model, repeating only the interrupted turn lowers EpG relative to resuming early and restarting late. The estimated savings over restarting early and resuming late are within the variation across tasks (Figure~\ref{fig:costs}d).
Rollback repeats only the interrupted turn, allowing us to test whether restoring more state than resume while repeating less work than restart lowers energy per successful goal.
On the main model, rollback had the lowest observed energy per successful goal of the eight recovery options after tool failures, both early and late (\hyperref[app:cells]{Appendix~\ref*{app:cells}}, \hyperref[tab:cells]{Table~\ref*{tab:cells}}).
For each type of late tool failure, warm rollback completes 27 of 30 goals correctly and cold rollback completes \qLateToolRbOk{} goals correctly.
Warm grounded resume completes \qLateToolSgOk{} goals correctly and warm resume completes \qLateToolSrOk{} goals correctly (Table~\ref{tab:glance} and \hyperref[app:cells]{Appendix~\ref*{app:cells}}, \hyperref[tab:cells]{Table~\ref*{tab:cells}}).
After early tool failures, warm rollback uses \rbFrEarlyTool{} percent less energy per successful goal than warm restart (95\% CI \rbFrEarlyToolLo{} to \rbFrEarlyToolHi{} percent) and \rbSrEarlyTool{} percent less energy per successful goal than warm resume (\rbSrEarlyToolLo{} to \rbSrEarlyToolHi{} percent).
After late tool failures, warm rollback uses \rbFrLateTool{} percent less energy per successful goal than warm restart (\rbFrLateToolLo{} to \rbFrLateToolHi{} percent) and \rbSrLateTool{} percent less energy per successful goal than warm resume (\rbSrLateToolLo{} to \rbSrLateToolHi{} percent).
Rollback ran in the second batch, where paired initial attempts used about \rbBatchRatioPct{} percent more energy.
Dividing second-batch energy by this ratio, assuming it also applies to recovery, preserves the observed minima and the direction of the resolved differences in this section (\hyperref[app:setup]{Appendix~\ref*{app:setup}}).

Rollback restores the conversation and tools to the start of the interrupted turn.
Resume keeps the error notice and restores tools only to just before the failed call.
The measured energy difference describes that combined operation.
After late tool failures, resuming completes 14 of 16 goals correctly for Mistral and 23 of 24 goals correctly for gpt-oss, and has lower energy per successful goal than the tested alternatives (\hyperref[app:cells]{Appendix~\ref*{app:cells}}, \hyperref[tab:m23]{Table~\ref*{tab:m23}}).

\paragraph{Choosing one fixed recovery.}
A framework using the same recovery for every failure needs to know the extra cost of that choice.
On the main model, warm rollback uses on average 6.5 percent more energy per successful goal than the cheapest recovery for each failure type and point (at most 18 percent, after late rate limits).
The mean excess cost is 10.1 percent for cold rollback, 15.2 percent for warm resume, and 18.7 percent for warm restart.
Warm rollback has the smallest mean excess cost among the fixed recoveries tested and repeats at most one turn of work (\hyperref[app:heldout]{Appendix~\ref*{app:heldout}}, \hyperref[tab:regret]{Table~\ref*{tab:regret}}).

\finding{2}{On the main model, repeating only the interrupted turn gives the lowest observed EpG after tool failures, both early and late. Warm rollback uses \rbSrEarlyTool{} percent less energy per successful goal than warm resume early and \rbFrLateTool{} percent less energy per successful goal than warm restart late, beyond the variation across tasks. The estimated savings over warm restart early and warm resume late are within the variation across tasks. Across all eight failure conditions, warm rollback has the smallest mean excess cost among the fixed recoveries tested (Figure~\ref{fig:costs}d, \hyperref[app:heldout]{Appendix~\ref*{app:heldout}}, \hyperref[tab:regret]{Table~\ref*{tab:regret}}).}

\subsection{RQ3: How Much Recovery Energy Cache Retention Saves}
\label{sec:cache}

\begin{table*}[!t]
\centering
\caption{Tokens and energy per successful goal for warm recoveries in the main study. Token counts are in thousands and include failed goals. Total counts input plus output, with cached input counted once as part of input. Unc.\ counts input tokens requiring fresh computation. EpG gives board energy per successful goal in kJ. Side identifies model-call or tool-call failures. Bold marks the lowest EpG in a row.}
\label{tab:tokens}
\vskip 0.1in
\footnotesize
\setlength{\tabcolsep}{5pt}
\begin{tabular}{@{}llrrrrrrrrr@{}}
\toprule
 & & \multicolumn{3}{c}{Restart} & \multicolumn{3}{c}{Resume} & \multicolumn{3}{c}{Rollback} \\
\cmidrule(lr){3-5}\cmidrule(lr){6-8}\cmidrule(lr){9-11}
Side & Point & Total & Unc. & EpG & Total & Unc. & EpG & Total & Unc. & EpG \\
\midrule
Model & early & 56.0 & 4.4 & \textbf{1.29} & 69.9 & 5.3 & 1.70 & 66.7 & 5.2 & 1.38 \\
Tool & early & 59.3 & 4.4 & 1.35 & 72.0 & 5.6 & 1.69 & 59.6 & 4.7 & \textbf{1.31} \\
Model & late & 82.1 & 4.7 & 1.86 & 56.4 & 4.4 & \textbf{1.29} & 63.4 & 4.8 & 1.52 \\
Tool & late & 80.7 & 4.6 & 1.82 & 65.1 & 5.2 & 1.51 & 61.3 & 4.6 & \textbf{1.45} \\
\bottomrule
\end{tabular}
\end{table*}

\begin{figure}[htbp]
\centering
\includegraphics[width=\columnwidth]{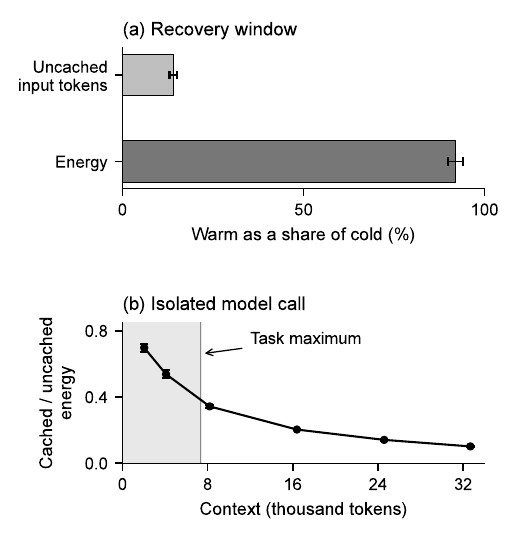}
\caption{\textbf{Cache retention saves a larger share of uncached input tokens than of recovery energy.}
(a)~Uncached prompt tokens and energy of a warm recovery as a share of the same recovery cold, main study, with 95\% bootstrap intervals over tasks.
(b)~Mean energy of an isolated cached call as a share of an uncached one, for Qwen3 30B on the same node, with 95\% bootstrap intervals over calls at each context size. Lower ratios mean larger savings. Shading marks contexts up to the task maximum.}
\label{fig:cache}
\end{figure}

At the task context lengths, cache retention removes most uncached prompt tokens but changes recovery energy much less (Figure~\ref{fig:cache}).
To measure the effect of a warm prefix cache, we compare warm and cold versions of the same recovery operation, allowing their subsequent calls and outcomes to differ (Section~\ref{sec:background}).
At task contexts up to \lawCtxMax{} tokens, a warm cache removes \cacheTokSave{} percent of a recovery's uncached prompt tokens (95\% CI \cacheTokSaveLo{} to \cacheTokSaveHi{} percent) but only \cacheEnergySave{} percent of its energy (\cacheEnergySaveLo{} to \cacheEnergySaveHi{} percent, Figure~\ref{fig:cache}a).

Uncached prompt tokens measure input that requires fresh prefill. Recovery energy also covers output generation, tool execution, and waits, so reusing prompt computation saves only part of that energy.
Cache retention nevertheless helps in most measured conditions. The cheapest recovery runs warm in six of the eight pairs of failure type and point, with the two cold winners within 2 percent of their warm versions.

The relative energy saving from caching grows with context in the isolated-call sweep (Figure~\ref{fig:cache}b).
On the same node, without tool time or waits, a cached call costs \sweepQwenCachedRatioFourK{} of the energy of an uncached call at 4k input tokens. The ratio falls to \sweepQwenCachedRatioLong{} near 32k input tokens.
Over the reported fits, uncached-call energy grows faster than linearly while cached-call energy grows more slowly (\hyperref[app:setup]{Appendix~\ref*{app:setup}}).

Tokens per successful goal divide the input and output tokens of all goals by the number that finish correctly, with cached input counted once.
Among warm restart, resume, and rollback, this measure selects the same cheapest recovery as energy in three of the four combinations of failure side and point. In the fourth, the recovery with the lowest EpG needs under 1 percent more tokens than the recovery with the fewest tokens.
After late tool failures, uncached-token counts favor restarting over resuming, while restarting has the highest energy per successful goal among these three warm recoveries (Table~\ref{tab:tokens}).

\finding{3}{At contexts up to \lawCtxMax{} tokens, keeping the cache warm removes \cacheTokSave{} percent of a recovery's uncached prompt tokens but only \cacheEnergySave{} percent of its energy. After late tool failures, warm restart uses fewer uncached tokens than warm resume but has higher EpG (Figure~\ref{fig:cache}, Table~\ref{tab:tokens}).}

\subsection{RQ4: Choosing a Recovery at the Failure}
\label{sec:choose}

\begin{figure}[htbp]
\centering
\includegraphics[width=\columnwidth]{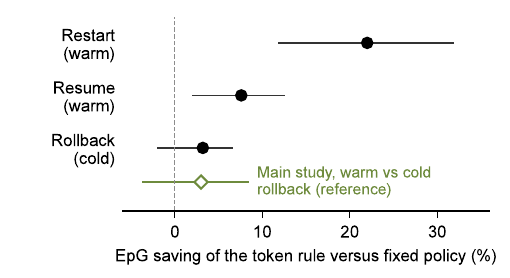}
\caption{\textbf{The token rule lowers EpG relative to always restarting and always resuming.}
Qwen3 30B, 27 complete held-out tasks, 108 goals per policy. The token rule selects the recovery predicted to need the fewest uncached prompt tokens. Points show percentage savings in pooled energy per successful goal over matched tasks. Bars are paired 95\% bootstrap intervals over tasks. Positive values favor the rule. The fixed rollback comparator is cold. The open diamond gives, for reference, the main-study saving of warm over cold rollback for the same failure types and points.}
\label{fig:choose}
\end{figure}

A rule fitted on earlier experiments lowers EpG relative to always resuming and always restarting on held-out tasks, beyond the variation across tasks (Figure~\ref{fig:choose}).
The token rule uses the failure type and number of calls already made to predict each option's total uncached prompt tokens during recovery, then chooses the smallest prediction before the outcome is known.
The token rule used \earVsResume{} percent less energy per successful goal than always resuming (95\% CI \earVsResumeLo{} to \earVsResumeHi{} percent).
It used \earSaveVsRestart{} percent less energy per successful goal than always restarting (\earSaveVsRestartLo{} to \earSaveVsRestartHi{} percent), with \earOkEar{} executed goals correct (Figure~\ref{fig:choose}).

The rule chose warm rollback in \trRbWarm{} of 111 decisions and warm resume in the remaining \trSrWarm{} decisions.
All resumes followed late failures, and \trSrWarmJson{} resumes followed malformed replies, where resuming is also cheapest in the main study.
Because most choices are warm rollback, a fixed warm rollback is the closest simple comparator.
The only rollback action executed as a fixed default on the held-out tasks was cold.
Against that default, the rule used \earVsBestFixed{} percent less energy per successful goal (95\% CI \earVsBestFixedLo{} to \earVsBestFixedHi{} percent, within the variation across tasks).
In the main study, warm rollback used \rbWarmColdHeld{} percent less energy per successful goal than cold rollback for the same failure types and points (\rbWarmColdHeldLo{} to \rbWarmColdHeldHi{} percent).
The rule's margin over cold rollback is close to this main-study saving, though the two estimates use different tasks. Executing fixed warm rollback on the held-out tasks would separate the benefit of choosing an operation from the benefit of retaining the cache.

\finding{4}{On held-out tasks, the token rule uses \earVsResumeInt{} percent less energy per successful goal than always resuming and \earSaveVsRestartInt{} percent less than always restarting, beyond the variation across tasks. It chose warm rollback in \trRbWarm{} of 111 decisions. Its estimated saving over always rolling back cold is within the variation across tasks (Figure~\ref{fig:choose}).}

\section{Discussion}
\label{sec:discussion}

\subsection{Choosing a Recovery in Practice}

To choose between restarting and resuming with Equation~\ref{eq:rule}, estimate the energy already spent, restart energy, success rates, and additional recovery energy on the intended tasks.

\paragraph{For agent frameworks.}
When the framework can restore both conversation and tool state, the main study favors warm rollback as a fixed recovery because it has the smallest mean excess cost among those tested (\S\ref{sec:drop}).
The recovery with the lowest energy per successful goal depends on the model, tasks, and failure, so a framework should measure again when these change.

\paragraph{For benchmark and measurement designers.}
Benchmark designers should report energy per successful goal beside recovery success rate, and distinguish the conversation and tool state restored by recovery from the prompt computation reused through the prefix cache.
They should also include failed goals and cached input when reporting tokens per successful goal (Section~\ref{sec:cache}).
Reports should state context lengths because the relative energy saving from cache retention grows with context.

\subsection{Scope and Limitations}

\paragraph{Failures and tasks.}
We study transient failures that reach the agent framework on multi-turn tasks the model solved reliably without injected failures.
Rollback requires tools whose state the framework can restore. Tools with irreversible external effects require safeguards that coordinate recovery with those effects~\cite{acrfence2026,wu2026saferesumebreakingexecution}.

\paragraph{Measurement setting.}
We measure one goal at a time at the board of one GB10 node, above its idle baseline.
Tool restoration, cache reset, and recovery decisions occur between measured attempt windows (Section~\ref{sec:method}). Slower restoration would add energy, which matters for close comparisons such as early rollback versus restart.

\paragraph{Models and serving.}
The comparisons describe recovery within each model's reliably solved tasks, and those task sets differ across models.
Concurrent serving changes cache reuse and requires a separate evaluation.

\paragraph{Next comparisons.}
Restoring only the conversation or only the tools would separate their roles in rollback, following AgentRewind's tests of context and environment restoration~\cite{agentrewind2026}.
Executing fixed warm rollback on the held-out tasks would show whether choosing an operation lowers EpG beyond retaining the cache.

\subsection{Reproducibility}
The decision library includes the calibration files and logged decisions, and replaying it on the recorded inputs reproduces all 111 logged token-rule choices~\cite{rafcode2026}.

\section{Conclusion}
\label{sec:conclusion}

Agent frameworks should compare recoveries by the energy needed for correct results, including energy spent on failed runs.
We compared eight recovery options after controlled call failures, with selected comparisons across three model families.
On the main model, restarting lowers energy per successful goal after early failures, while resuming lowers it after late failures.
Restarting's higher success rate and the additional energy used during resumption account for the early saving. Late in a task, preserving the work already completed makes resuming cheaper.
Among the fixed recoveries tested on the main model, warm rollback has the smallest mean excess cost.
At our task context lengths, a warm prefix cache saves a much larger share of uncached prompt tokens than of recovery energy.
We suggest that framework developers choose how their agents recover by measuring energy per successful goal on their intended models and tasks, and that benchmarks report this cost beside recovery success rate.
The cost of a correct result includes the energy of the runs that fail.

\FloatBarrier
\providecommand{\doi}[1]{doi: \href{https://doi.org/#1}{\nolinkurl{#1}}}
\bibliography{ear_master}
\bibliographystyle{mlsys2025}
\clearpage
\appendix
\section*{Appendix}
\setcounter{dbltopnumber}{2}
\renewcommand{\theHsection}{app.\Alph{section}}
\renewcommand{\theHtable}{app.\arabic{table}}
\renewcommand{\theHfigure}{app.\arabic{figure}}
\renewcommand{\theHequation}{app.\arabic{equation}}

\section{Measurement and Validity}
\label{app:setup}

The energy comparisons in Section~\ref{sec:method} use the same measurement boundary across recoveries within each study and retain goals whose logs confirm the planned failure.
A goal is one task run, including its initial execution and any recovery.
Energy per successful goal (EpG) divides total measured goal energy, including failed goals, by the number that finish correctly.
A warm prefix cache retains computed prompt state for reuse, while a cold cache clears it.

\paragraph{Energy boundary.}
The board's DC-input rail covers the whole node, including memory, storage, fans, and conversion losses outside the chip.
We report board energy above the idle baseline measured with the model loaded and no requests.
The package counter covers the chip, while NVIDIA Data Center GPU Manager (DCGM) reports the GPU alone.
We read board and package power at 10~Hz and GPU power at 1~Hz, and check the ordering of these readings for each run.
As a consistency check, we compare the sensor readings and find board energy above package energy and package energy above GPU energy.
The earlier Qwen3 4B pilot study predates the board reading and uses the chip's package counter.
In the main study, the time between attempt windows is at most 1.4~s.
Every study ran vLLM 0.23.0 with prefix caching and chunked prefill on, at most 4 concurrent sequences and a maximum model length of 32{,}768 tokens.

\paragraph{Study coverage.}
Table~\ref{tab:studies} distinguishes the main comparison from studies that vary decoding, the model, or the kind of error.
Sampling tests whether the ordering recurs when the model draws different replies, and the additional models use the tasks each solves reliably.
Wrong-step runs test errors that the tool accepts, so the agent receives no error notice.

\begin{table}[t]
\centering
\caption{Study configurations. All run on one GB10 node, one goal at a time. Goals include failure-free reference runs, and the Mistral and gpt-oss goals include failure-free screening of all 30 main-study tasks.}
\label{tab:studies}
\vskip 0.1in
\footnotesize
\setlength{\tabcolsep}{2.6pt}
\begin{tabular}{@{}llr>{\raggedright\arraybackslash}p{1.55cm}r@{}}
\toprule
Study & Model & Tasks & Variation & Goals \\
\midrule
Main & Qwen3 30B & 30 & 4 failure types, 8 options & 2{,}010 \\
Sampling & Qwen3 30B & 10 & temperature 0.7, 3 seeds & 240 \\
Wrong steps & Qwen3 30B & 26 & accepted wrong step & 239 \\
Held-out & Qwen3 30B & 28 & rules and fixed recoveries & 666 \\
Mistral & Mistral 24B & 8 & 3 recoveries & 252 \\
gpt-oss & gpt-oss 20B & 12 & 3 recoveries & 348 \\
Pilot & Qwen3 4B & 10 & restart and resume & 350 \\
\bottomrule
\end{tabular}
\end{table}

\paragraph{Injection and validity.}
The injector counts model calls and tool calls separately within the first attempt.
Let $n$ be the number of calls of the applicable kind in the first failure-free screening run.
An early failure fires at the second call of that kind. A late failure fires at the rounded value of $0.7n$, clamped to $[3,n-1]$.
The task configuration stores the resulting call numbers for model-side and tool-side failures.
A wrong step replaces an argument of a tool call and counts only if the tool accepts it.
A valid goal has exactly one firing, in the first attempt, at the planned failure type, position, and call number. An invalid goal is rerun once.
The accepted-wrong-step summary additionally excludes cases marked self-corrected. Its early denominators range from 10 to 15 across recovery options, although the study inventory contains 26 tasks in total.
The warm resume and warm restart counts reported in Section~\ref{sec:point} each have 13 retained goals, with membership determined separately for each recovery.
The counts describe those retained populations. A causal comparison of retaining the wrong step would require matching the task populations as well.

\paragraph{Shared history and the second batch.}
To assess how closely the experiments approximate a common state at failure, we check the histories of the initial attempts compared in Section~\ref{sec:method}.
We compare the first attempts call by call, prompt and reply, after replacing generated call and request identifiers.
The four options within each batch have identical calls and replies in \sameHistMain{} of 240 combinations of task, failure type, and point in the first batch and in \sameHistRb{} of 240 combinations in the second.
Across batches, warm restart and warm resume each match warm rollback up to the failed call in \sameHistCross{} of 240 combinations.
Across the 240 shared combinations, the second batch's initial attempts measured \rbBatchRatioPct{} percent more board energy than the first batch's.
They ran about 2 percent longer, and the idle power subtracted for that batch was 0.5 to 0.8~W lower.
As a sensitivity check, we divide second-batch energy by the pooled initial-attempt energy ratio.
The observed winners and the direction of the resolved paired differences in Section~\ref{sec:drop} remain the same under this adjustment, which assumes that the ratio also applies to recovery attempts.

\paragraph{Paired confidence intervals.}
The intervals in Section~\ref{sec:method} use 2{,}000 draws with replacement from the tasks shared by every recovery and failure condition in a comparison.
Each draw uses the same tasks for both recoveries and all included failure conditions. A draw with zero successful goals for either recovery has undefined EpG and is omitted from the interval calculation.
The resulting intervals describe variation across the selected tasks.

\paragraph{Isolated calls at long context.}
Figure~\ref{fig:cache}b uses the Qwen3 30B sweep on the GB10 node, with \sweepQwenUncachedNCalls{} uncached and \sweepQwenCachedNCalls{} cached calls across six context sizes up to \sweepQwenCachedCtxMax{} tokens.
At each size, the ratio divides mean cached-call energy by mean uncached-call energy. The fits and ratios use 95\% intervals from 1{,}000 resamples of calls with replacement, independently within each cache state and context size.
The reported uncached fit has a positive quadratic coefficient, $\sweepQwenUncachedQuadD{}$~J/token$^2$ (95\% CI $\sweepQwenUncachedQuadDLo{}$ to $\sweepQwenUncachedQuadDHi{}$).
Cached-call energy grows more slowly, with a fitted linear slope of \sweepQwenCachedLinB{}~J/token. This difference accompanies the decreasing cached-to-uncached ratio in the figure.
These single-call measurements extend beyond the task contexts used for the recovery comparison.

A separate A100 sweep measures GPU energy only. Its positive quadratic coefficient is \sweepQuadD{}~J per (1k tokens)$^2$ (95\% CI \sweepQuadDLo{} to \sweepQuadDHi{}).
That result supports the qualitative context dependence, but its hardware and measurement boundary differ from the GB10 board measurements. The task-level recovery comparisons of Section~\ref{sec:cache} use contexts only up to \lawCtxMax{} tokens.

\section{Results by Failure Type and Point}
\label{app:cells}

The tables below support the recovery comparisons reported by model-call and tool-call failures in Section~\ref{sec:results}.
Table~\ref{tab:cells} gives the main study's warm recoveries and cold rollback, including the lowest observed EpG in each failure condition. Table~\ref{tab:m23} gives the comparisons on the additional model families.
\begin{table*}[!t]
\centering
\caption{Main study (Qwen3 30B, 30 tasks). The ok columns count correct goals. J/run gives mean energy per run, including failed runs, and EpG gives energy per successful goal, both in board joules. Bold marks the lowest observed EpG of all eight recovery options in a row. The displayed warm recoveries and cold rollback include the lowest option in each row.}
\label{tab:cells}
\vskip 0.1in
\footnotesize
\setlength{\tabcolsep}{3.4pt}
\begin{tabular}{@{}llccrcccrccccc@{}}
\toprule
 & & \multicolumn{3}{c}{Restart, warm} & \multicolumn{3}{c}{Resume, warm} & \multicolumn{2}{c}{Rollback, warm} & \multicolumn{2}{c}{Rollback, cold} & \multicolumn{2}{c}{Grounded, warm} \\
\cmidrule(lr){3-5}\cmidrule(lr){6-8}\cmidrule(lr){9-10}\cmidrule(lr){11-12}\cmidrule(lr){13-14}
Type & Point & ok & J/run & EpG & ok & J/run & EpG & ok & EpG & ok & EpG & ok & EpG \\
\midrule
Malformed reply & early & 28/30 & 1{,}231 & \textbf{1{,}319} & 22/30 & 1{,}403 & 1{,}914 & 23/30 & 1{,}437 & 22/30 & 1{,}641 & 23/30 & 1{,}754 \\
Rate limit & early & 28/30 & 1{,}173 & \textbf{1{,}257} & 25/30 & 1{,}264 & 1{,}517 & 23/30 & 1{,}332 & 22/30 & 1{,}526 & 25/30 & 1{,}482 \\
Tool error & early & 28/30 & 1{,}267 & 1{,}357 & 22/30 & 1{,}248 & 1{,}702 & 26/30 & \textbf{1{,}314} & 26/30 & 1{,}367 & 25/30 & 1{,}589 \\
Timeout & early & 28/30 & 1{,}258 & 1{,}347 & 22/30 & 1{,}237 & 1{,}686 & 26/30 & \textbf{1{,}312} & 26/30 & 1{,}382 & 25/30 & 1{,}587 \\
\midrule
Malformed reply & late & 28/30 & 1{,}784 & 1{,}912 & 28/30 & 1{,}245 & \textbf{1{,}334} & 26/30 & 1{,}561 & 28/30 & 1{,}527 & 27/30 & 1{,}468 \\
Rate limit & late & 28/30 & 1{,}691 & 1{,}812 & 28/30 & 1{,}161 & \textbf{1{,}244} & 26/30 & 1{,}471 & 28/30 & 1{,}465 & 27/30 & 1{,}354 \\
Tool error & late & 28/30 & 1{,}711 & 1{,}833 & 24/30 & 1{,}208 & 1{,}510 & 27/30 & 1{,}439 & 29/30 & \textbf{1{,}421} & 26/30 & 1{,}543 \\
Timeout & late & 28/30 & 1{,}692 & 1{,}812 & 24/30 & 1{,}207 & 1{,}508 & 27/30 & 1{,}471 & 29/30 & \textbf{1{,}447} & 26/30 & 1{,}585 \\
\bottomrule
\end{tabular}
\end{table*}

Warm restart and warm resume have the same early--late ordering for time per successful goal as for energy.
After early failures, restart takes \timeEpgFrWarmEarly{}~s per successful goal and resume takes \timeEpgSrWarmEarly{}~s per successful goal. After late failures, resume takes \timeEpgSrWarmLate{}~s per successful goal and restart takes \timeEpgFrWarmLate{}~s per successful goal.
In the token comparison (Section~\ref{sec:cache}), cached input accounts for \cachedSharePct{} to 94 percent of prompt tokens, so including it changes how tokens rank recoveries after late failures.

\begin{table}[!htbp]
\centering
\caption{Mistral Small 3.2 24B (8 tasks) and gpt-oss 20B (12 tasks), each on the tasks it solves reliably. The ok columns count correct goals, and EpG gives board joules per successful goal for warm restart, warm resume, and cold rollback. Bold marks the lowest observed EpG in a row.}
\label{tab:m23}
\vskip 0.1in
\footnotesize
\setlength{\tabcolsep}{2.4pt}
\begin{tabular}{@{}llcrcrcr@{}}
\toprule
 & & \multicolumn{2}{c}{Restart} & \multicolumn{2}{c}{Resume} & \multicolumn{2}{c}{Rollback} \\
\cmidrule(lr){3-4}\cmidrule(lr){5-6}\cmidrule(lr){7-8}
Type & Point & ok & EpG & ok & EpG & ok & EpG \\
\midrule
\csname @@input\endcsname tables/m23_table_clean
\end{tabular}
\end{table}

\section{The Cost Model}
\label{app:derivation}

The condition in Section~\ref{sec:metric} compares energy per successful goal from a common state at failure.
Let $S$ be the energy already spent at failure, $E$ the expected energy of a restart recovery, and $p_R>0$ and $p_K>0$ the success probabilities of restarting and resuming.
The simplified cost model estimates unfinished work by $E-S$ and writes resumption energy as $E-S+\Delta$.
Thus $\Delta$ is the residual above this estimate.
The costs of restarting and resuming are
\begin{align*}
J_R &= \frac{S+E}{p_R}, &
J_K &= \frac{S+(E-S)+\Delta}{p_K}
     = \frac{E+\Delta}{p_K}.
\end{align*}
Multiplying $J_R<J_K$ by the positive product $p_Rp_K$, we obtain $p_K(S+E)<p_R(E+\Delta)$, which rearranges to the condition in Equation~\ref{eq:rule},
\[
S < \left(\frac{p_R}{p_K}-1\right)E+\frac{p_R}{p_K}\Delta.
\]
When $p_R=p_K$, the restart recovery energy $E$ cancels and the condition becomes $S<\Delta$.

\begin{samepage}
For rollback, let $T$ be the energy of the completed portion of the interrupted turn that must be repeated and let $p_B>0$ be its success probability.
Assuming the remaining work otherwise has estimated energy $E-S$, total expected goal energy is $E+T$.
With positive expected goal energy when resuming, $E+\Delta>0$, we obtain
\[
J_B=\frac{E+T}{p_B}<J_K
\quad\Longleftrightarrow\quad
\frac{p_B}{p_K}>\frac{E+T}{E+\Delta}.
\]
\end{samepage}
If rollback and restart have equal success probability, $p_B=p_R$, rollback is cheaper than restarting when $T<S$.
Rollback then repeats less completed work while preserving the same chance of success.
These conditions compare recoveries once their energies and success rates are supplied. Choosing from call counts requires calibrating those quantities.

The terms reported in Section~\ref{sec:point} are measured per failure type and point from the warm restart and warm resume goals of the main study.
We estimate the common $S$ by averaging the mean board energies of the initial attempts under restart and resume.
The mean energy of a restart's recovery attempt gives $E$, and the observed success rates give $p_R$ and $p_K$.
We obtain $\Delta$ by subtracting $E-S$ from the mean energy of a resume's recovery attempt.
The residual is positive for each tested failure type and point, so resumed execution uses more energy than this estimate of the remaining work.
Substituting these measured terms reproduces the observed ordering when both recoveries use this shared estimate of $S$. This calculation accounts for recorded costs and outcomes. Prediction requires estimates available when the failure occurs.

\section{Held-Out Study}
\label{app:heldout}

Table~\ref{tab:e3} supports the comparison in Section~\ref{sec:choose} between rules that choose a recovery during execution and fixed recovery choices.

\begin{table}[!htbp]
\centering
\caption{Held-out study (Qwen3 30B, 28 tasks, 111 decisions per policy). The ok column counts correct goals, J/run gives mean energy per run including failed runs, and EpG gives energy per successful goal, both in board joules. The paired savings use 27 complete tasks (108 goals per policy), so they differ from ratios of these all-goal EpGs. Bold marks the lowest observed EpG, whose paired difference from always rollback (cold) has an interval that includes zero (Section~\ref{sec:choose}).}
\label{tab:e3}
\vskip 0.1in
\footnotesize
\setlength{\tabcolsep}{4pt}
\begin{tabular}{@{}lccc@{}}
\toprule
Policy & ok & J/run & EpG \\
\midrule
Learning controller & 95/111 & 1{,}248 & 1{,}459 \\
Threshold rule & 98/111 & 1{,}242 & 1{,}406 \\
Token rule & 100/111 & 1{,}167 & \textbf{1{,}296} \\
\midrule
Always restart (warm) & 101/111 & 1{,}512 & 1{,}662 \\
Always resume (warm) & 97/111 & 1{,}236 & 1{,}414 \\
Always rollback (cold) & 101/111 & 1{,}231 & 1{,}353 \\
\bottomrule
\end{tabular}
\end{table}

The token fits use the main study and earlier experiments on wrong steps and failures that persist during recovery. Their tasks are disjoint from the held-out set.
Each fit predicts uncached prompt tokens during recovery as an intercept plus a coefficient times the total model and tool calls made before failure.
In the released decision records, the token rule chose warm rollback in \trRbWarm{} of 111 decisions and warm resume in \trSrWarm{}~\cite{rafcode2026}.
The \trSrWarm{} resumes followed late failures, including \trSrWarmJson{} malformed replies and one tool error.
The predictions placed warm rollback below warm resume in \trRbWarm{} decisions, with a median gap of \trPredGap{} uncached tokens across all decisions.
The fitted predictions determine these choices for individual tasks and failures.
A fixed warm-rollback choice would agree with the rule in 98 cases. Measuring its energy and correctness on these tasks requires executing that fixed choice as well.

The learning controller represents uncertainty in success with a Beta distribution for each failure type and recovery, then draws one success rate for each recovery before choosing.
It scores each recovery by its predicted recovery energy divided by that sampled rate. Equation~\ref{eq:J} also includes energy spent before failure, which affects comparisons when success rates differ.
Below its restart threshold, it considered only restart options in 56 of its 111 recorded decisions.
Its fitted parameters came from the calibration preceding the correction to how repeated goals were matched across studies.
The energy predictions also used zero token counts. Supplying the recorded token counts in a replay diagnostic changes \ctrlContextChangedDecisionsN{} choices at the first recovery attempt.
These settings define the learning-controller configuration evaluated here.
The threshold rule used \xstarRuleOnlineVsFr{} percent less energy per successful goal than always restarting.
The learning controller used \ethreeCtrlVsFrWarmSave{} percent less energy per successful goal than always restarting.
Table~\ref{tab:regret} supports the fixed-recovery comparison in Section~\ref{sec:drop}, giving each fixed recovery's excess EpG relative to the cheapest option for each failure type and point, identified after observing outcomes.
Because that reference is selected from the same data, it gives an optimistic comparison.
The released library includes a replay test that reproduces the 111 recorded token-rule choices~\cite{rafcode2026}. Evaluating their energy and correctness requires the experimental records alongside those choices.

\paragraph{Threshold calibration.}
The thresholds supplied to the rule range from 3.9 to 8.1 calls across failure types and retain the earlier calibration.
The held-out study measures how that rule performs during execution. Validating the call thresholds against Equation~\ref{eq:rule} requires fitting them to its energy and success-rate terms.

\begin{table}[!htbp]
\centering
\caption{Extra energy per successful goal of a fixed recovery relative to the cheapest recovery for each failure type and point, identified after observing outcomes (percent, board). The first three columns summarize each study's measured recoveries. The held-out column gives measured executions relative to the best of its three fixed recovery choices. Entries average over eight failure conditions in the main and additional-model studies and four in the held-out study. A dash marks an untested recovery, and bold marks the smallest excess on Qwen3 30B.}
\label{tab:regret}
\vskip 0.1in
\footnotesize
\setlength{\tabcolsep}{3.5pt}
\begin{tabular}{@{}lcccc@{}}
\toprule
Policy & Qwen3 30B & Mistral & gpt-oss & Held-out \\
\midrule
Restart (warm) & 18.7 & 27.3 & 36.7 & 25.5 \\
Resume (warm) & 15.2 & 8.7 & 23.8 & 6.8 \\
Rollback (warm) & \textbf{6.5} & -- & -- & -- \\
Rollback (cold) & 10.1 & 13.6 & 32.8 & 2.2 \\
\bottomrule
\end{tabular}
\end{table}

\end{document}